# Towards responsible research in digital technology for health care

Pierre JANNIN, INSERM, University of Rennes, Rennes, France

**Abstract**

Digital technology is everywhere for the benefit of our daily and professional life. It strongly impacts our life and was crucial to maintain professional and social activities during the COVID19 crisis. Similarly, digital technologies are key within biomedical engineering research topics. Innovations have been generated and introduced over the last 40 years, demonstrating how computing and digital technologies have impacted health care. Although the benefits of digital technology are obvious now, we are at the convergence of several issues which makes us aware about social, societal and environmental challenges associated with this technology. In the social domain, digital technologies raise concern about exclusion (financial, geographical, educational, demographical, racial, gender, language, and disabled related exclusion) and physical and mental health. In the societal dimension, digital technologies raise concern about politics and democracy (sovereignty and governance, cognitive filters and citizen's engagement), privacy and security (data acquisition and usage transparency, level of personal approval, and level of anonymization), and economics. In the environmental dimension, digital technologies raise concern about energy consumption and hardware production. This paper introduces and defines these challenges for digital technology in general, as well as when applied to health care. The objective of this paper is to make the research community more aware about the challenges of digital technology and to promote more transparency for innovative and responsible research.

## 1 Introduction

Computers are everywhere, benefitting our daily and professional life. We use them for writing, reading, communicating, learning, at school, at home, at work, and in our cars. We use them for healthcare, administration, government, finance, weather forecasts, sports, music, arts, and science. Especially during crises, including the current health crisis, computers are crucial to keep professional and social activities possible, offering easy mechanisms to communicate, meet, and interact with each other. Computers and **digital technologies** have become central to our life and our world and, undoubtedly, they have been useful and beneficial. They are key to the technological progress we have followed the last 150 years and have strongly impacted our world. Digital technologies also revolutionized medicine and improved understanding of anatomy, physiology and pathology. Digital technologies have enabled the introduction of digital medical imaging and computer-assisted interventions, for better diagnosis, therapy and evaluation.

This digital revolution is probably the fastest innovation we have ever had in our history. Such breakthrough innovation contributes to the *technological progress*. However, philosophers and sociologists, as well as voices from the citizens, are starting to question and challenge the notion progress in the sole term of technological progress. This is a positive sign of an adult and mature society. According to Kranzberg's six laws, "technology is neither good nor bad; nor is it neutral." (1) The French historian and sociologist Jacques Ellul said that it is important to consider that a technology includes all good and bad uses, and is concerned with more than just how humankind uses it. Etienne Klein, a French scientist and sociologist, explained that there has been a trend in the past towards reducing the definition of the progress to technological progress and especially its economic impact. The notion of the "*Anthropocene*" was proposed in 2000 (2,3) to define the times when the global environment, at some level, started to be shaped by human activity rather than vice versa. This was part of a global consciousness about the relation between our technological progress and the world in which we are living. The definition of the progress has evolved throughout history (from the Latin word *progressus*: the action of moving forward). Considering the maturity of our today's world, E. Kein considers that **progress** should be defined as an evolution of society and humankind towards better understanding, and respect for our planet, and increased happiness. Voices are emerging from respected national and international institutions to consider this global vision of progress to shape tomorrow's world and the corresponding actions and challenges (4).

In this debate, it is important to consider that the discussion is about technology and not about the underlying science. Science generates new knowledge and explores new domains through observation and experimentation. **Technology** is the application of this knowledge and these domains for various specific purposes and it is this technology that needs to be studied. The issue is therefore to question technology as it relates to the holistic definition of progress: to question its contribution to progress including both its positive and negative aspects; to promote good and virtuous practices and to minimize and control bad and negative ones; to ensure that technology is useful AND responsible; and to adapt the technology to society, and not the inverse. It is also important to start considering responsible digital technologies from the very first Technology Readiness Levels (5). Technological research is not only in the hands and responsibility of industry but also of research. We do both technological and fundamental research applied to medicine and health care, with a strong influence of the targeted clinical applications and impacts, regularly going until prototype development and evaluation. Obviously, a new technology has strengths, weaknesses and may induce opportunities and threats (SWOT methodology (6)). All these aspects need to be considered when assessing a technology and reported when proposing a new technology.

Of course, the purpose of this paper is not to oppose the pros and cons of technology. The question is not a battle between one vision and its inverse, between fans and opponents of technology. It is the search for discussion and debate. The purpose is neither to take a stand between utilitarianism and egalitarianism, but to ask for systemic (including ethical) analysis. Since strengths and opportunities of a technology are usually well covered in publications, the emphasis is given here in identification of weaknesses and possible risks (i.e., threats).

With the rise of Machine Learning (ML), and deep learning in particular, there have been a recent awareness about the ethical impact of research (7,8,9). In (7), the authors introduced 8 principles for Responsible Machine Learning.

- **Human augmentation** refers to the need to assess the impact of incorrect predictions and to design systems cognizant of the role of the human user (such as human-in-the-loop review processes);
- **Bias evaluation** refers to the need for understand, document, and monitor bias during development, training, testing, validation, and production;
- **Explainability by justification** outlines the need to improve the transparency and explainability of ML systems where reasonable;
- **Reproducible operations** refer to the need for reproducibility in ML, displacement strategy to the need to mitigate the impact towards workers being automated, practical accuracy to the need to ensure that accuracy and cost metric functions are aligned to the domain-specific applications;
- **Trust by privacy** outlines the need to protect and handle data with stakeholders that may interact with the system directly and/or indirectly; and finally

- **Data risk awareness** outlines the need to ensure data and model security throughout the ML lifestyle including training, dissemination, and algorithm deployment.

The Montreal Declaration for responsible Artificial Intelligence development goes a bit further (8) with 10 principles. Most relate to the ethical development and use of Artificial Intelligence (AI) systems: to increase of quality of life and well-being, solidarity, with respect of privacy and intimacy, equity, democracy and diversity. It also includes the principle of prudence and responsibility for the development and use of AI. The last concerns sustainable development and environmental issues. In (9), the authors reviewed the primary documents and frameworks that have been proposed for ethical use of AI and scored each in terms of main challenges, specifically privacy, accountability, safety and security, transparency and explainability, human control of technology, professional liability, promotion of human values and international human rights.

Challenges related to AI and machine learning applied to health care were mainly addressed with regards to possible methodological biases: such as those resulting from inappropriate learning data sets, those related to inappropriate statistical distributions, and those resulting from wrong, inadequate or inappropriate validation methodology. In (10), the authors showed that gender imbalance in two standard image datasets used to diagnose various thoracic diseases resulted in consistent decrease in performance of three deep neural network architectures for women. In (11), the authors implemented a deep learning hip fracture prediction system that uses X-ray radiographs. They showed that when associated with various patient and hospital process related variables, the system performance can be boosted, with the most relevant variable was the scanner brand name. There are still few analyses of technological challenges in digital technology for health care regarding social, societal and environmental dimensions.

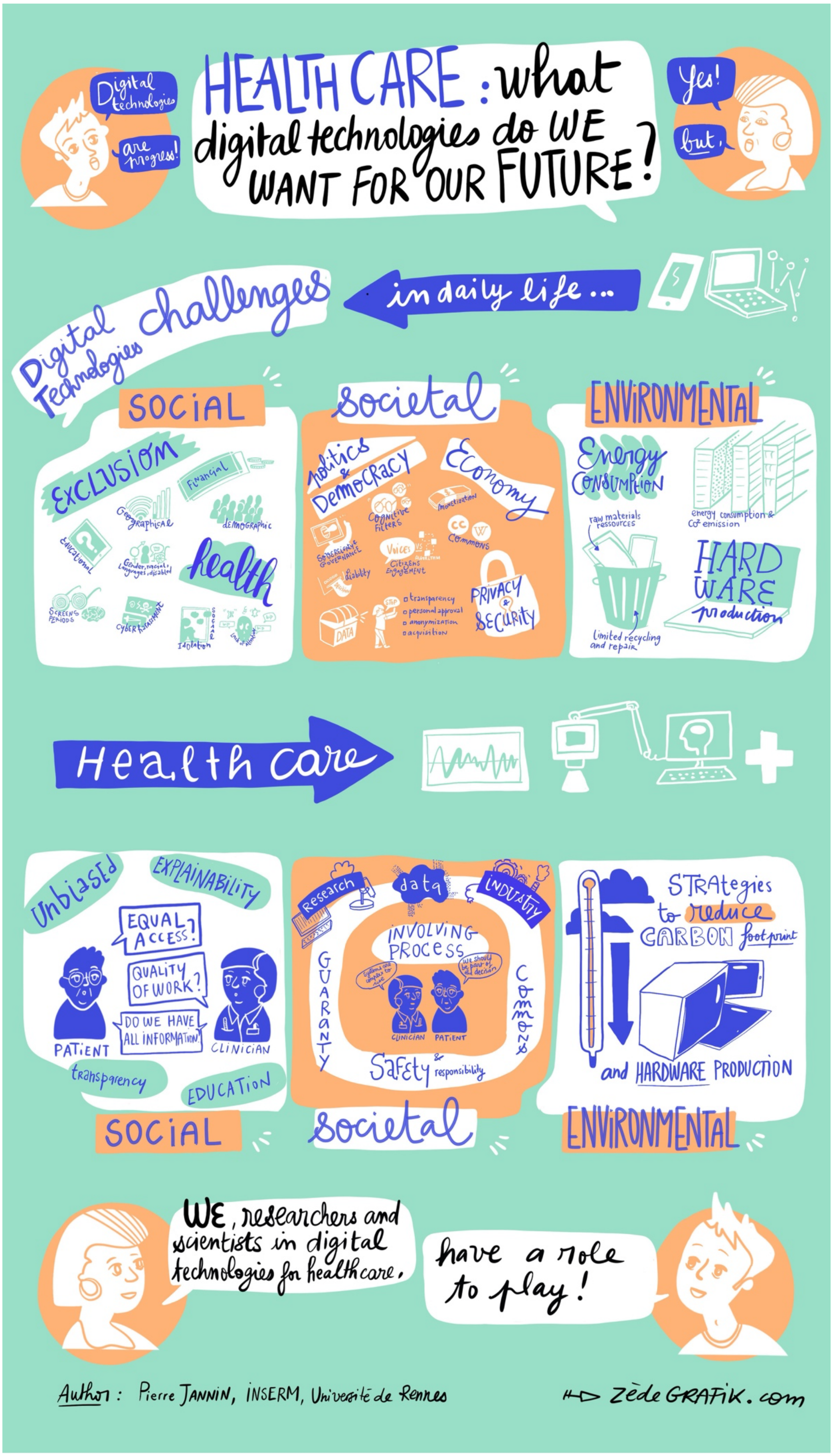

Figure 1: Challenges in Digital Technology in daily life and in health care: Towards Responsible research

The purpose of this paper is to address the notion of **responsible technological progress in digital technology for health care** specifically in computer assistance for diagnosis and interventions. I will analyze digital technology by exploring the notion of responsible research divided into three dimensions: social, societal and environmental (Figure 1). The social dimension addresses the challenges related to individuals. The societal dimension studies challenges related to collective society. The environmental dimension studies challenges related to the local and global environment. For each dimension, I will first list the challenges related to digital technology in general and then deduce from that the corresponding challenges related to the digital technology for health care. The objective of this paper is to make our community more aware about the challenges of digital technology and to promote more transparency for innovative and responsible research.

# 2 Social dimension

## 2.1 Impacts of digital technologies on individuals

We have already emphasized the uses and benefits of digital technologies for individuals in both personal and professional life. This has been especially remarkable during the COVID19 crisis where working from home has become a standard driving the use of videoconference systems for meeting both colleagues and family. This trend towards the digitization of every piece of data and information to facilitate computing, analysis, representation, communication and prediction is not likely to stop. It will probably cover all aspects of our professional and personal life soon. This is also why the analysis of weaknesses and potential risks is crucial.

The weaknesses and potential risks of digital technology from a social point of view are becoming more well identified and quantified. Why is it recent? It is probably because until recently we did not have enough history, nor enough data about them. Possibly, the pressure towards technology development has pushed economic and financial issues to the forefront, hiding other parameters.

Focusing on its weaknesses and potential risks, the digital technology social dimension includes two main challenges: **exclusion and health** (Figure 2).

**Exclusion** is defined as the situation in which some individuals do not benefit from digital technology. There are many reasons that may explain this exclusion, which include *financial, geographical, educational, demographic, racial, gender, language, and disability-related exclusion*.

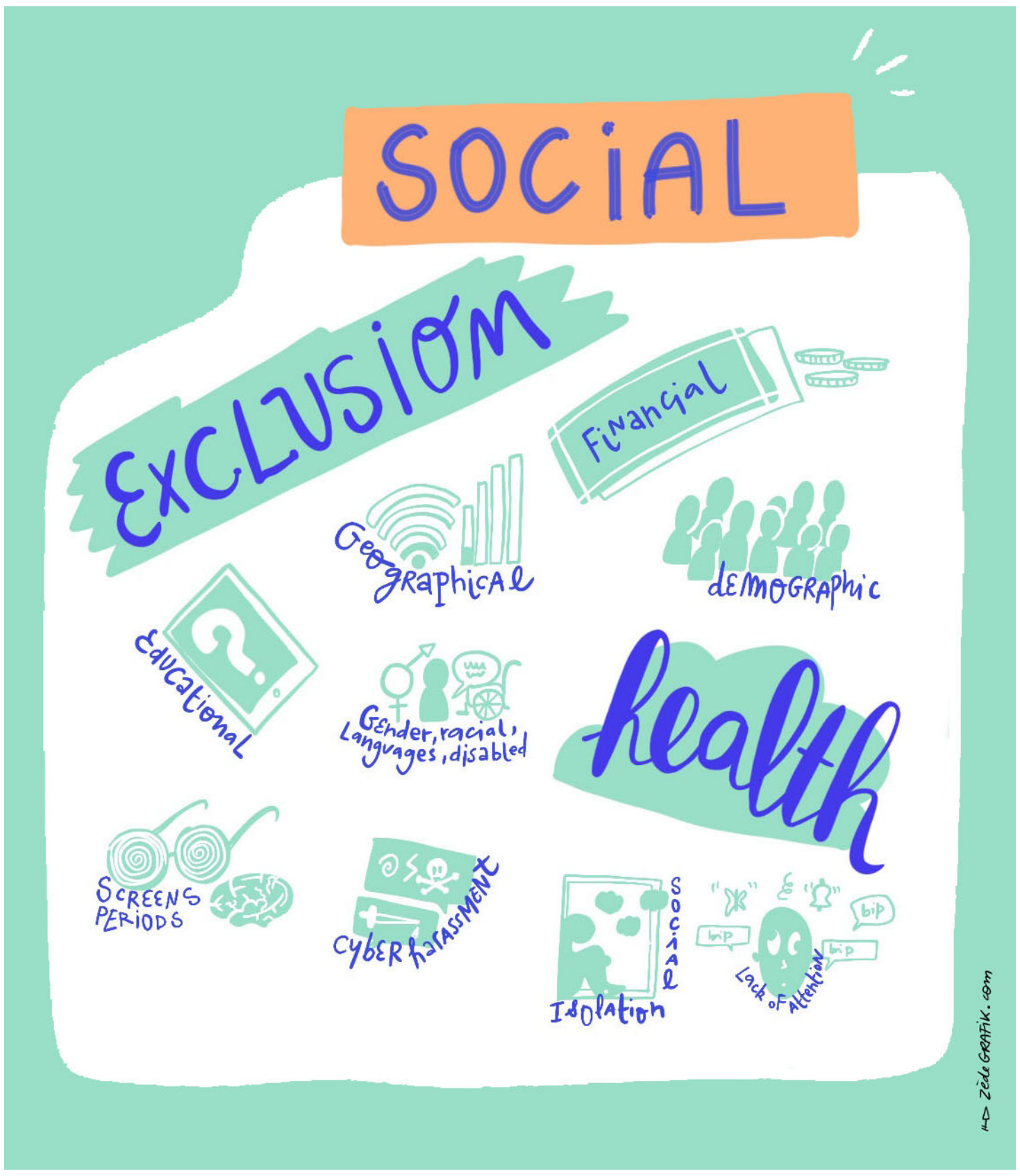

*Figure 2 : Social challenges in Digital Technology in daily life*

*Financial exclusion* is mainly related to the global cost of digital technologies including purchase, maintenance and use costs. While hardware (e.g., computers and cell phones) are becoming cheaper, they are still expensive for low-income individuals. Whereas, in some European countries, an internet subscription costs less than 10 USD per month, it costs more than 100 USD in some African countries and around 60 USD in North America. Knowing that several subscriptions may be needed for a single family, the budget for digital technologies is around 250USD per month in most western countries. This digital technologies related budget has been shown to be a concern for 14% of adults in the United States during the past few months of the COVID-19 outbreak (13), significantly influencing remote education in poorer populations. All in all, there is an important relationship between poverty and low level of digital technologies adoption (14).

*Geographical exclusion* is related to financial exclusion but not exclusively so. Some territories do not have access to wireless telecommunications, or high-speed fiber optics. There is a clear geographical exclusion related to developing countries. Whereas northern European countries have a level of internet penetration around 97% in 2021, many countries are left behind, especially in Africa where the level of internet penetration is around 25% (12). In 2019, almost half of private households worldwide were estimated to have a home computer, spanning from 80% in developed countries to 30% in developing ones (17). These disparities also exist within a country itself. The rate of having a wired high-speed internet connection at home varies from 93 to 71% across the different states in the US as of November 2019 (15).

*Educational exclusion* refers to the lack of education about digital technology relative to its use, availability and risks. Digital technology introduced a specific way of communicating with systems, with specific human/machine interfaces (e.g., mouse, cursor, menus, checkboxes). Wide digitization makes computer and software usage mandatory. Not only does software have different modes for user interaction but also different types of devices may be used ranging from personal computers to tablet PCs and cellphones. Depending on the operating system and hardware brand, user interfaces may also differ. In addition, communication between humans and computers is changing continuously as new interface devices and human-machine interaction paradigms are developed. Many segments of the population have never learned the logic of digital technologies, let alone programming. Until very recently, general education about digital technologies was not really considered as a scholarly activity and thus not widely addressed in usual educational contexts. A recent national study in France showed that more 39% of the population felt uncomfortable completing administrative processes on-line; 37% found that the administrative processes became more complex with digitization (16). In addition, education when available rarely includes education about both usage and weaknesses and possible risks, on social, societal and environmental dimensions, with suggestion of good practices.

Additional factors increase the risks of exclusion: *demographic, racial, gender, language, and disability-related exclusions.* These exclusions are, in a certain sense, double: exclusion from access to digital technologies combined with direct or indirect bias in digital technologies, especially in recent machine and deep learning approaches. Digital technology is a "white man's world" (18) especially in commercial settings. Only 25% of "computer and mathematical occupations" workers were female in US in 2020 and only 9% and 8% of digital technologies workforce were African-American or Hispanic, respectively (18). In addition, few user interfaces and digital technologies are adapted to diversity. For instance, few user interfaces and digital technologies are adapted to people with disabilities. Digital technologies are not adapted to minorities yet. This has been epitomised in Natural Language Processing, where gender and racial biases due to the (under-)representativity of data sets has led to four different categories of impact: 1) denigration - use of culturally or historically derogatory terms-, 2) stereotyping - reinforcing existing societal stereotypes, 3) recognition inaccuracy, and 3) under-representation - disproportionately minimizing a specific group (20). These biases resulting from unrepresentative datasets have direct impact on the performance of medical machine learning algorithms as well (21). The well-known story of Microsoft's AI twitter account TAY that was closed after less than 24 hours of functioning illustrates how uncontrolled AI systems with algorithms fed with inappropriate data may turn into a racist and far right extremist chat-bot (33).

Exclusion in digital technologies is also about by purpose implementation of inequity in algorithms. For instance, the algorithm behind the Tinder application, which aims to optimize matching between peoples to facilitate social connections, is based on the optimization of a function. The cost function, also called a desirability score, is based on personal data (such as age, gender, and level of education computed by quality of writing analysis), a learning factor from favorites profiles (learned from swipe actions performed by the user) and additional rules. One of these rules increases the score for men with higher level of education whereas it decreases for women (22). Then, the optimization strategy is inspired by the Chess scoring ELO system where peoples with same desirability scores are matched (23,24). An Austrian employment agency developed a software that scored job seekers to evaluate their probability to find a job. Job seekers with higher scores had given higher priority. This score was automatically decreased for non-Austrian, women, women with children, peoples over 30, and disabled-peoples. The European data protection authority (GDPR) has prohibited the use of the system from January 1, 2021 (25,26).

Several studies demonstrated how inappropriate use of digital technologies may impact **health**, both physiological and psychological. Long periods in front of screens may not only have vision but also neurological impacts. A French study showed that 1 of 5 school students are subjected to cyber-harassment (27) and another has shown that the excessive use of social networks results in greater social isolation and a degradation of self-esteem (28). The way the Internet is built now focuses on the value of attention, which is captured through three main approaches: retention, alerts, and immersion. Many cues are implemented to capture attention including dark patterns and gamification, resulting in addictive behavior such as binge watching (29).

## 2.2 Corresponding challenges for digital technology for health care

It is unarguable that digital technologies have deeply changed medicine for the patient's benefit, leading to better clinical outcomes, earlier treatment for known diseases, new treatments for diseases that were previously unaddressed. We are at the forefront of this technological revolution and this is still the beginning. However, digital technology for health care is subject to the same social challenges than digital technologies in general.

*Exclusion*: The first aspect is to ensure as much as possible that all peoples can similarly benefit from the developed technology, minimizing financial, geographical, educational, demographic, racial, gender, language, and disabled-related exclusions. If this cannot be done for a particular study, researchers should report this. If it is done by purpose, by design (e.g., a solution for a dedicated therapy, a rare disease, a specific population or an objective of positive discrimination), this has to be reported too. The quest for egality and equity should be at least considered with regards to the basic needs of the humanity, such as the ones defined by the UN SDGs (4). Regarding the cost of technology, both costs of production and usage should be addressed or justified and initiatives in cost reduction and affordable or frugal technology should be promoted (such as the FAIR MICCAI 2021 workshop and the ISCAS frugal technology award). Developers should be aware about usage-related exclusion and should privilege technology that is easy to use for a broader range of the population with particular attention to ergonomics and human factors. Intrinsic exclusion resulting from inappropriate machine learning strategies should be controlled, reported, and avoided as much as possible and rigorous bias-conscious validation schemes should be adopted. The use of a new technology should include education for both medical users and patients. Being able to understand the basics of the technology involved in the digital system, including its performance and its limitations, is crucial. Professionals must be familiar with the use of the system, which should be easy to use, unambiguous, and easy to understand. Patients have also to know about the system and the involved technology, their possible bias and uncertainties, as well as the logics that was followed to propose a decision. There is an emerging need from some citizens to be informed, especially when some parts of the decision-making process are delegated to digital technology. This will be also addressed below with regards to societal related challenges. This also implies that researchers and developers should integrate transparency and explainability into their algorithms.

*Health*: The second aspect is to ensure that new developed technologies do not cause harm. Obviously, in our domain, our core motivation is to improve patient health. It is becoming clear that this notion of improvement of health is adopting a wider definition of health that includes quality of life and not only clinical outcomes. The health challenge also includes the professionals' health: clinicians, radiologists, surgeons, general doctors, scrub nurses, and anesthesiologists, for instance. Impacts on their quality of work (fatigue and posture, for instance), as well possible psychological impacts of delegated decisions and actions for healthcare professionals, have to be addressed. Loss of autonomy, loss of skills, and overtrust are potential human consequences on professionals that should be considered with care.

# 3 Societal dimension

## 3.1 Impacts of digital technologies on society

Digital technology has offered several important positive impacts for society as a whole. The most well-understood is economic, which has primary focused on productivity through process optimization that enables institutions and commercial enterprises to do more with lower costs. This productivity growth generated new business, new jobs, and new wealth as well as revolutionizing all sectors of the economy. However, the weaknesses and potential risks of digital technologies from a societal point of view have also started to be identified. This societal dimension includes **political and democratic, privacy and security, and economical** challenges.

While the **political and democratic** challenges of digital technologies seem less obvious than the others, there are still extremely important. “Algorithms are political » said D. Cardon, a French sociologist (30). Political and democratic sub-challenges include *sovereignty and governance*, *liability*, *cognitive filters,* and *citizen engagement* (Figure 3).

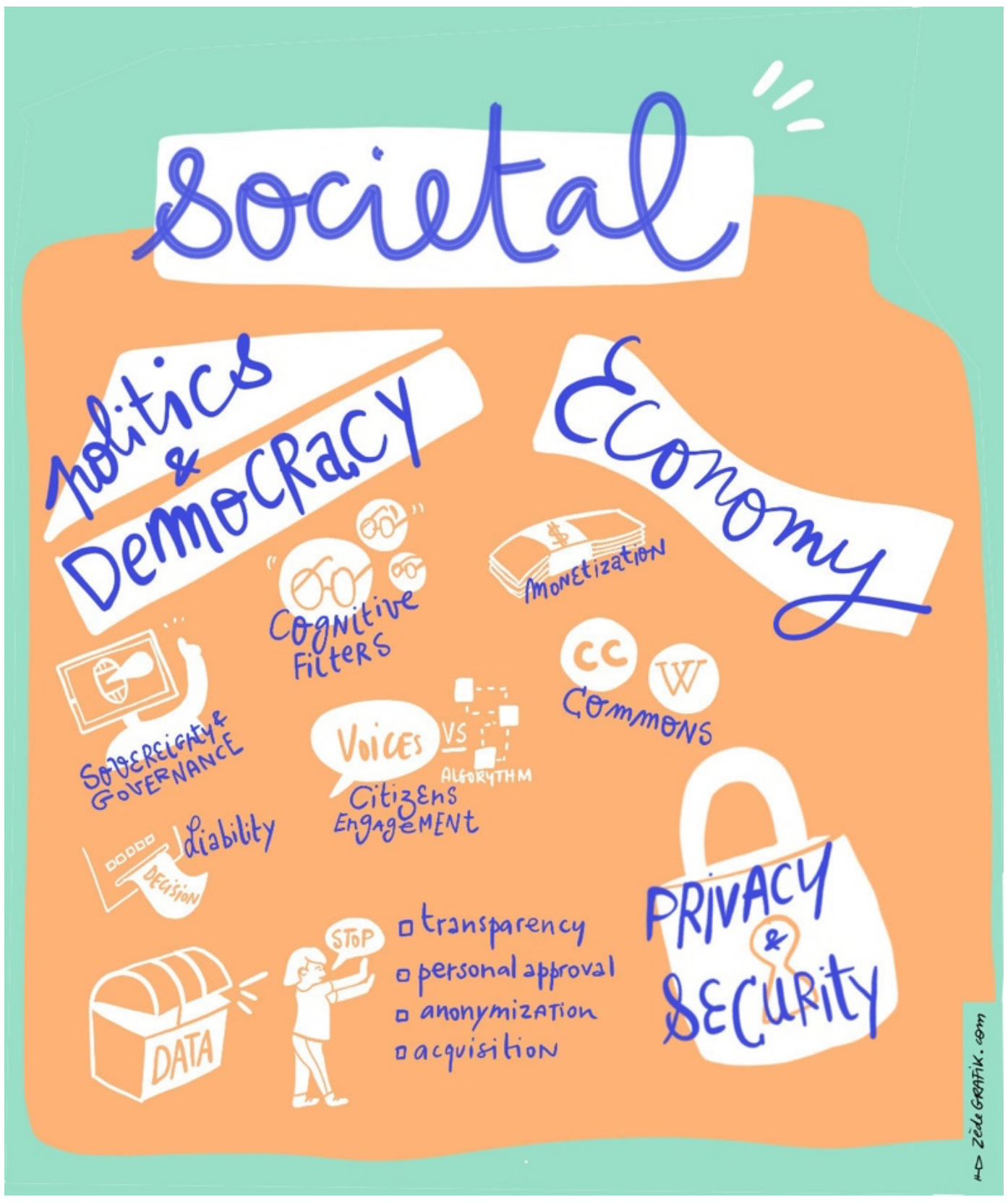

Figure 3: Societal challenges in Digital Technology in daily life

*Sovereignty and governance* are challenged by the introduction of decision-making algorithmic assistance in economy, law, and governance. Sometimes, a decision-making role is given to algorithms which are not always explainable, especially when using deep learning. There is a critical risk for democracy involved in delegating certain decisions to algorithms. Explainability, defined responsibility, safety management, and human cooperation are mandatory in the usage of AI systems. In the European General Data Protection Regulation (GDPR), Articles 13–15 mention the rights for the citizen that benefit from (or subject to) an automated decision to have access to "meaningful information about the logic involved" in the automated decision. This is a first step towards the right to explanation (31) and transparency requirement.

The internet, which is a key element of digital technologies, is not neutral either. Software, hardware, and operating system choices are never neutral and thus impact control and sovereignty. The main regulatory institution managing and creating domain names and extensions is the ICANN. It is a non profit organization but fully independent and disconnected from governments. Internet contents is now in the hands of the main digital giants, including search engines and social medias. Public institutions, such as governments, do not directly regulate the internet. While this was an initial choice for more liberty on the net, a shift in responsibility may now be warranted to counterbalance the almost unbounded power of a limited number of actors.

As mentioned above, there is a critical risk for the democracy in delegating certain decisions to algorithms. Thus, *citizen engagement* in the decision-making process is mandatory. This involvement has three main objectives: 1) to counterbalance the growing role of automation in decision making, 2) to reinforce trust in technologies, science, and public institutions, and 3) to drive innovation by use rather than by technology. Involving citizens in political decision making has been deeply studied and rigorous methodologies have been proposed to minimize bias in the recruiting and consulting process. The amount of involvement of citizens spans from no role to full control (34). Whatever the level of engagement, participation is conditioned on an important initial education stage where citizens meet experts and are made aware about technical information and related challenges. At the no-role extremum, citizens are informed about the decision, its context, the reasons for the decision, and the corresponding action. At the other extremum, these decisions and actions are fully implemented by the citizenry. In between, consulting adds a stage where citizens propose decisions and actions. This could include a post-processing stage in which citizens are made aware about what the implemented decisions and actions were as well as their results and evaluation; with some approaches having the possibility for citizens to follow-up in a long-term iterative loop.

*Liability* is also challenged when decisions and actions are performed by algorithms or automation systems. There is much debate regarding the needs for clear regulation regarding liability whatever the level of autonomy taken by the digital technologies (from shared decision and action between human and machine to full autonomous systems). Dedicated laws and regulations are needed in (semi-) automated decisions with AI and (semi-) automated actions in robotics. The distribution of liability between researchers, industry, and users is unclear yet.

Social networks have been identified as a major source of "*cognitive filtering*", creating "cognitive bubbles" or "echo chambers". Cognitive filtering is the process of automatic and algorithmic selection of information, news, and media which are derived from your favorites, your previous selections, and your usual requests and digital activities potentially on other platforms (32). Through cognitive filters, you are progressively being restricted your digital world. Cognitive bubbles are closed small and focused communities that reinvent their own vision of the world, their own truth and are created from cognitive filters. It is generally associated with a lack of trust in classical media, science, politics, and public institutions and identified of one of the possible contexts explaining the Capitol attack in Washington DC in January 6, 2021.

Both **security and privacy** issues are important in improving trust in digital technologies by citizens and users and consequently their trust in science, in public institutions, and in politicians. It is mandatory to carefully consider trust to avoid technology rejection as a whole. Both issues are the most cited challenges as regards to machine learning and artificial intelligence (35).

Whereas a lot (if not all) of personal administrative and financial processes are digitally managed, **security** is crucial for both individual and public interests. Digital security is becoming crucial also to avoid misuse, whereas connected systems (such as the internet-of-things (IoT)) are more easily hacked. If digital technologies are used for decision making in politics, law, or health care, ensuring protection of such systems is imperative. Resiliency of systems should also be guaranteed to ensure continuity of services. It is important to implement mechanisms for ensuring and demonstrating trust in digital technologies. Different categories of mechanism are possible such as public (e.g., open-source implementation), private (e.g., multi-factor authentication, password strength requirements) and digital (e.g., blockchain) ones. Here, the choice of mechanism is again a political one.

Digital technologies and especially AI technology strongly rely on data availability. This induces several challenges about respect of individual data **privacy**. This privacy aspect includes challenges about *data acquisition and usage transparency*, *ownership*, *level of personal approval*, and *level of anonymization*. *Data acquisition and usage transparency* is crucial. Whereas European regulations within the General Data Protection Regulation (GDPR) provide a more or less clear approval process for personal data collection and usage in the internet, this is not true for all countries or all data acquisition systems. For instance, personal gas and electricity meters are now equipped with direct wireless connections to servers that collect that data. Sensors are installed in some cities to capture traffic or air quality. Some are basic sensors with few dedicated measurements and a low temporal resolution whereas others rely on video. Such sensors are not always visible to, or known by, citizens. How all these different data types are used in the short and long term is not always clear. Several international cases are well known where data collected with a particular intent was sold for a different usage, indirectly transferred or captured by private companies (36). ***Ownership*** **of data, information and knowledge is therefore an important question.** It is also. Data processing may generate information and knowledge, extracting additional value from raw data. Clear conditions of ownership for both raw and derived data, information and knowledge are key challenges related to both the economic and political (sovereignty and governance) dimensions. The *level of approval* by the user for any acquisition of personal data is also an additional issue. There are still discussions around a mandatory approval by the user or just an informative message to the user. In addition, the *level of anonymization* is important to state and explain. Is it de-identification, pseudonymization, or real anonymization? Is full anonymization warranted or even feasible? Surely, data privacy relies on trust and digital trust, clear regulatory and governance schemes, and clear economic models, which should all play a part in the digital commons (see below).

The *model* of the **economic** dimension of digital technologies is not the one which was envisioned 30 years ago and will probably not be the same 30 years from now. For the internet, we recently moved from what was called a data economy to an information and knowledge economy. More recently, the economics of attention emerged from the importance of social networks (as defined by H. Simon, Nobel Prize 1978 in economic sciences). Being able to keep individuals under the control of patterns (see above) that capture their attention also allows for them to be redirected to specific information, websites, or even habits. Another facet of this is the economy of opinions which looks for expressions freely given by individuals to score websites, comment on news, rate restaurants, or like videos. Whereas the concept of the "wisdom of the crowd" motivated these free reviews, they have been regularly demonstrated as inaccurate, and strongly biased. Opposing of the monetization of data, information and knowledge, there is a tendency toward of the concept of the "digital commons" or "wikinomics" which too was found in the earlier days of the internet (37). The perfect example of this is Wikipedia which was based on collective intelligence and has even been considered as good and accurate as world renowned Encyclopedia Britannica (38). Wikinomics is based on well-organized and reviewed collaboration and cooperation to generate freely available data, information, and knowledge. In analogy to a common, *digital commons* are digital resources which should be accessible to all humans (39). These resources are held in common, not owned privately and should be carefully protected by digital trust strategies. These commons include open-source software, wikis, and open data, protected by specific legal frameworks such as the Creative Commons or the GPL licensing agreements. Although these resources are currently legal or digital abstraction, if a move towards a more integrated digital society is desired, hardware and connectivity resources such as computing hardware, basic internet access, high speed wired connection with public optical fiber networks, should also be considered.

### 3.2 Corresponding challenges for digital technology for health care

Similarly, digital technology for health care also faces societal challenges.

*Politics and democracy*: The first aspect relies on the relationship between the patients and the clinicians with the developed technology and the corresponding system. When digital technologies are used to improve decision-making processes, it is crucial to involve clinical organisations, public bodies, and patient representatives in a governance scheme to ensure the quality and relevance of the developed technologies. This will become even more relevant with the advance of artificial intelligence. Similarly, liability is a sensitive issue in health care if digital technologies are allowed to be part of the medical decision-making and action processes. In addition, involving both end users and patients, in the design, evaluation, and also choice of clinical problem is also a way to work towards responsible research in digital technology for health care. Some countries have worked towards this by adding expectations in grant proposals regarding patient engagement. Finally, transparency and explainability are still more relevant when automated decisions impact quality of life and health integrity of patients.

*Security and privacy*: The importance of security and safety in medical devices, software, and systems in general is still extremely relevant in our area and there is probably insufficient research effort put towards these aspects. It is our responsibility to at least identify possible weaknesses and security breaches within our systems. Data privacy issues are also very important in health care and especially in machine learning. Similar issues regarding data acquisition and usage transparency, level of personal approval, and level of anonymization are well known by our community and stakeholders and are the subject of regulations worldwide. However, there is still no consensus regarding how to navigate the trade-off between data availability and privacy. There is a need to address and discuss these issues involving established stakeholders from industry, hospitals, patient representatives, governments, and our own research communities. Access to clinical data with outcomes is, without any doubt, extremely valuable for learning best practices but without clear and convincing explanation, without trust, and with growing cyber-security threats, some patients may not be willing to share their personal data.

*Economy*: While the digital healthcare economy is important worldwide and should continue to contribute to the progress of healthcare for the benefit of patients and professionals, it is also important to foster a higher level of access for the maximum number of patients. For this reason, it is also important to consider the concept of digital commons. This is something that our community has started to understand, with the promotion of open data and challenges (40), and the growing number of open-source software. But there is still an important effort to make by broadening such initiatives, extending it to medical digital commons, and open computing resources, for instance.

# 4 Environmental dimension

## 4.1 Impacts of digital technologies on the environment

The main advantage of digital technologies for the environment rests on the possibility of monitoring and optimizing energy and resource consumption by mean of sensors, IoT technology, wireless communications, and artificial intelligence. This is the combination that usually underlies smart cities or smart industries. This reduction of energy consumption is real but limited and has been estimated to be about 20% when smart city software is implemented (41), and about 15% for smart home software (42). However, the environmental impact of digital technologies is becoming very sensitive due to the intersection of two facts: 1) the impact of energy consumption and human activities on climate change is now well admitted and 2) digital technology is growing significantly (44).

The Sixth Assessment Report from the Intergovernmental Panel on Climate Change (IPCC) was released on August 2021 (43) and its conclusions are clear. There is a clear link between global warning and climate change: "*With every increment of global warming, changes get larger in regional mean temperature, precipitation and soil moisture.*" Human activities impact the recent global warming and climate change: "*It is unequivocal that human influence has warmed the atmosphere.*" Solutions require "*limiting cumulative CO2 emissions, reaching at least net zero CO2 emissions, along with strong reductions in other greenhouse gas emissions (GHG).*" Under different simulated "*scenarios with low or very low greenhouse gas, discernible differences in trends of global surface temperature would begin to emerge from natural variability within around 20 years, and over longer time periods for many other climatic impact-drivers*."

The environmental and energy impact of digital technology is due to both **energy consumption** and **hardware production** (Figure 4).

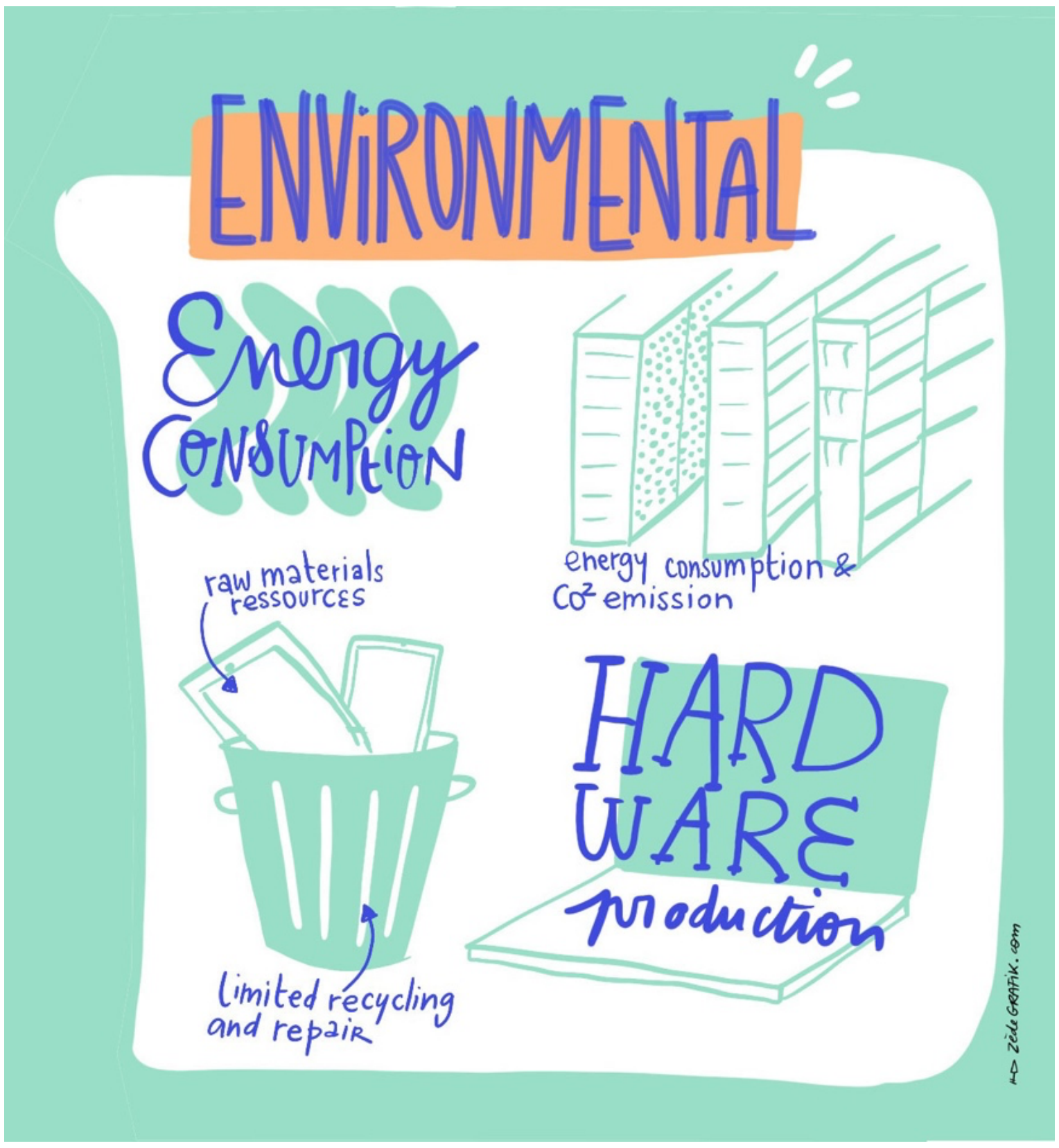

*Figure 4 : Environmental challenges in Digital Technology in daily life*

With the growth of digital technologies, its **energy consumption** will correspondingly grow, consumption both during *use* and *conception*. Digital technology represents between 6 and 10% of the global electricity consumption. Its energy consumption is due both to usage (55%) and production of hardware (45%). It is responsible of about 4% of greenhouse gas emissions, whereas civil aviation is about 2% and cars about 8% (45). Its CO2 emission is expected to significantly grow in correlation to the exponential growth of digital technologies (it is impossible to give reliable figures due to the diversity of values available in the literature). Its GHG emission is due to personal use (47%) versus network infrastructures and data centers (53%) (46). In terms of internet use, video streaming takes 60% of the bandwidth in 2019 (47) and is expected to grow to 80% in 2022 (48), with a corresponding growth in energy consumption. Reducing humanity's carbon footprint is an effort of which individuals in their daily life should be responsible for 25%, and collectives or organizations, industry in particular, should be responsible for 75% (49). Software conception is also energy consuming and it is becoming more critical with artificial intelligence (54). Learning a deep network consumes a large amount of energy. Despite what we could innocently imagine, algorithmic and coding strategies may have strong impact on energy consumption and consequently on CO2 emissions (55,56).

The **production of digital technologies hardware** is also a challenge by itself. It includes issues about amount of required *raw material resources* and limited *recycling and repair*. The production of a 2kg laptop requires about 800kg of raw materials. It also requires rare-earth metals, the extraction of which strongly impacts the environment and the health of miners (50,51), with some of these metals being exhausted in as little as 50 years (52). Digital technologies hardware is not only about computers. It also includes all the connected equipment from IoT technology which is expected to multiply by a factor of 3 to 5 by 2025 (53). In terms of the hardware production issue, there is little repair or reuse (less than 1%), as well as little recycling of digital technologies hardware (about 35%), mainly due to the growing complexity and planned obsolescence of both hardware and software (57,58). In France, less than half of Waste Electrical and Electronic Equipment (WEEE) is collected; only 18% is collected worldwide (42% in Europe, 12% in Asia, 10% in America and 1% in Africa) (59). In addition, the average lifespan of a computer went from 11 years in 1985 to 4 years in 2015. Additionally, transport of materials as well as resulted digital technologies has also an important impact on energy consumption and GHG emission; knowing that digital economy also strongly modified the relationship not only between production location and selling location but also between selling point and consumer location and also increased transport of goods.

Advances in digital technologies on both computing and transmission could be envisaged as a way to reduce energy consumption, but it does not consider the possible so-called "rebound effect": the more you have, the more you use, canceling out the expected improvement (60). To use an analogy, with a car twice as fast, you are more likely to go twice as far, thus spending the same amount of time in the car despite its improvement in speed. Most of these environmental issues are based on diagnoses which are difficult to scientifically validate. However, it outlines a tendency about a definitely non-null environmental impact of digital technology and allows an initial awareness about environmental issues in digital technology.

### 4.2 Corresponding challenges for digital technology for health care

Since we strongly rely on digital technology, our domain faces the same environmental challenges. However, there are few initiatives, in our domain, addressing these issues.

*Energy consumption*: The energy consumption and the corresponding carbon footprint of the proposed methods or systems should also be reported and controlled (61). There are now tools that exist to do this estimation such as the Green Algorithms calculator (62), and Carbontracker (63). Similarly, main literature about energy consumption in machine learning, despite their discrepancies regarding absolute numbers, agreed about the need to estimate, report and control energy consumption and carbon footprint (55,56). Here again, some tools exist, such as the Machine Learning Emissions Calculator (64).

*Hardware production*: There are currently no numbers about this, but there is generally little attention towards optimizing hardware production in our area. There is, to my best knowledge, no existing recycling process or secondhand market within our community. There is no repair café set up between researchers or academic community. However, as it was mentioned above, they are some initiatives that encourage development of low-cost technology and affordable solutions by dedicated workshops or dedicated awards. There are also initiatives such as hackathons, where students and researchers meet together to develop, during few days, solutions and prototypes in an agile approach, and fablabs, where citizens, whatever they are scientists or not, work together in dedicated spaces to develop low-cost solutions for well-defined needs (65). These initiatives should be supported and developed.

## 5 Discussion

As researchers and scientists, we are trying to develop innovative concepts and systems that help improve the life of patients. We also try to increase the knowledge of humanity about itself and its environment. For this, we rely on a strong scientific methodology that helps to develop innovative and relevant solutions. And it is our responsibility as a researchers and developers of future technologies to rigorously follow the best methodologies.

Over the last 30 years, we have benefited from a large effort, not only in developing new technologies, but also in validating and evaluating these technologies (66,67,68,69). Nowadays, validation and evaluation of a technology are seen as part of its research and development process. Validation is necessary for first studying the output of a method or a system. Following this, demonstration of its added value through evaluation studies is also crucial. This evaluation component assesses the value of a new method in the context of its use and usually involves user studies (68). In this paper, I propose to extend such evaluation methodology to include an assessment of the broader impacts of the proposed technology, introduced in the previous sections. Similar to previous validation and evaluation methodologies (67,69), this extension requires frameworks, methodologies, and tools to evaluate the impacts regarding three dimensions: social, societal, and environmental. The proposed framework relies on the challenges and sub-challenges identified in the previous sections (Table 1).

Indeed, two stages could be followed towards responsible research: 1) to evaluate, estimate and report impacts of the proposed technology and then 2) if relevant, to optimize retrospectively (or, even better, prospectively) some impacts of the technology, with regards to performance and usefulness. Being able to report both in parallel may help identifying weaknesses and understanding relative added value of the solution. Stage 1 should rely as much as possible on existing standardized metrics, tools, and methodologies for the qualitative and quantitative estimation of the impacts. For stage 2, there are also existing methodologies that may facilitate the optimization of some challenges. Some have been mentioned above, for instance, for optimizing energy consumption for machine learning methods. There are also best practices that can easily be considered and may have a direct impact. Privileging low-cost solutions and open science with open licensing schemes, facilitating recycling, reuse, and repair are approaches that both positively impact these social, societal, and environmental dimensions. Involving both end users and patients in the design, evaluation, and also the choice of clinical problem is also a way to work towards more responsible research in digital technology for health care. Of course, a technology will probably never be able to fulfill all challenges and possible impacts: being inclusive, low-cost and low carbon foot print, … Some are by design limited to a dedicated population or a rare disease. The quality and the relevance of a technology and its implementation should be assessed as regards the tradeoff between strengths, weaknesses, opportunities, and threats.

As scientists and researchers developing new healthcare technologies, we have the responsibility to be aware about the many facets in which technology impacts humankind, and not exclusively about its scientific rigor and quality, nor its innovative aspect only. As it was mentioned in the introduction, the objective of this paper is not at all to slow down technological innovation. However, for all of us, time and resources are limited, choices need to be made, priorities need to be set.

| Challenges | Sub-challenges |
|---|---|
| Social dimension | |
| Exclusion | *Financial* |
| | *Geographical* |
| | *Educational* |
| | *Demographic* |
| | *Racial* |
| | *Gender* |
| | *Language* |
| | *Ability* |
| Health | *Physical* |
| | *Mental* |
| Societal dimension | |
| Political and democratic challenges | *Sovereignty and governance* |
| | *Citizen engagement* |
| | *Liability* |
| | *Cognitive filtering* |
| Security and Privacy | *Security* |
| | *Data acquisition transparency* |
| | *Data usage transparency* |
| | *Data, information & knowledge ownership* |
| | *Level of personal approval* |
| | *Level of anonymization* |
| Economical | *Models* |
| | *Commons* |
| Environmental dimension | |
| Energy consumption | *In use* |
| | *In conception* |
| Hardware production | *Raw material resources* |
| | *Recycling and repair* |

*Table 1: Challenges and sub-challenges for responsible research and development in digital technology for health care*